\documentclass[twocolumn,showpacs,amsmath,amssymb,prl]{revtex4}

\usepackage{graphicx}
\usepackage{dcolumn}
\usepackage{bm}

\def\kem{\tilde{\kappa}_{e-}}

\def\kop{\tilde{\kappa}_{o+}}

\def\b1{\beta_{1}}
\def\b2{\beta_{2}}
\def\b3{\beta_{3}}

\begin{document}

\title{Rotating Odd-Parity Lorentz Invariance Test in Electrodynamics}

\author{Michael E. Tobar$^{1}$, Eugene N. Ivanov$^{1}$, Paul L. Stanwix$^{2}$, Jean-Michel G. le Floch$^{1}$,John G. Hartnett$^{1}$} 
\affiliation{$^{1}$School of Physics, University of Western Australia,  $^{2}$Harvard-Smithsonian Center for Astrophysics}

\date{\today}

\begin{abstract}
We report the first operation of a rotating odd-parity Lorentz Invariance test in electrodynamics using a microwave Mach-Zehnder interferometer with permeable material in one arm. The experiment sets a direct bound to $ \kappa_{tr}$ of  $-0.3\pm 3\times10^{-7}$. Using new power recycled waveguide interferometer techniques (with the highest spectral resolution ever achieved of  $2\times10^{-11} rad/\sqrt{Hz}$) we show an improvement of several orders of magnitude is attainable in the future. 
\end{abstract}

\pacs{07.60.Ly, 06.20.-f, 03.30.+p, 07.50.Hp, 84.40.-x}

\maketitle

\section{I. Introduction}

The Standard Model Extension (SME) extracts terms from additional Lorentz and CPT violating fields that preserve the local gauge symmetry of the usual standard model of particle physics. Thus, in the photon sector the SME leads naturally to an extension of Quantum Electrodynamics with extra Lorentz and CPT violating photon fields \cite{Kosto1_1, Kosto1_2, Kosto1_3}. This work focuses on renormalizable components of the SME in the photon sector as first described in \cite{KM, Kosto}, which involves operators of mass dimension four or less. Recent developments by the same authors allowing for operators of arbitrary mass dimension have not been considered in this work \cite{KM_arb}. Of the nineteen independent components, ten associated with vacuum birefringence have been constrained by astrophysical tests to no more than parts in $10^{37}$ \cite{KM2006}. This leaves nine components, including the scalar component $\kappa_{tr}$, the $3\times 3$ symmetric traceless $\kem^{jk}$ matrix with five degrees of freedom, and the $3\times 3$ antisymmetric $\kop^{jk}$ matrix with three degrees of freedom.

Even parity experiments have leading order sensitivity to the $\kem^{jk}$ Lorentz violating coefficients. Examples of these are the modern Michelson Morley experiments \cite{Lipa, Stanwix, Stanwix2, Tobar2, Herrmann, Mueller:2007, Antonini, Peters09, Schiller09}, which have set limits on these coefficients of order $10^{-17}$. The odd parity coefficients $\kop^{jk}$ are only sensitive through the boost of the experiment with respect to the considered frame of reference (which in this case is a sun centered frame). This means the sensitivity is suppressed by the value of the boost, which in this case is of order of the Earth orbital speed ($10^{-4}$). The sensitivity to $\kappa_{tr}$ should be second order sensitive to boost for these type of experiments, even though the details are yet to be fully calculated one expects the sensitivity to be reduced by a factor of the boost squared with a suppression factor of order $10^{-8}$.

Examples of odd parity experiments have been recently shown to be of the Ives-Stilwell type \cite{Tobar, Reinhardt, Hoh, Saathoff, Bailey, MP07, Carone}. These type of experiments are leading order sensitive to the odd parity coefficients $\kop^{jk}$, with boost suppression sensitivity to $\kappa_{tr}$. Thus, these experiments have been used to set limits to the least well known parameter $\kappa_{tr}$, since Michelson Morley experiments have set limits on the even and odd parity coefficients of order $10^{-17}$ and $10^{-13}$ respectively. In this work we take this approach and perform the first rotating odd parity experiment to set a limit on $ \kappa_{tr}$.

It may seem counterintuitive that a rotating experiment is sensitive to the isotropic Lorentz violating coefficient, however because the odd parity experiment is leading order boost sensitive to $\kappa_{tr}$, a rotating experiment will detect a time varying signal as it changes its orientation with respect to the vector boost. Thus, for a constant rotating experiment, a signal will be expected at the rotation frequency for a non-zero value of $\kappa_{tr}$. However, due to the sidereal rotation of the Earth with respect to the sun centered frame and the orbital motion around the sun, mixing between these frequencies causes signals at sidereal and annual frequency offsets from the rotation frequency, which places the frequencies of interest away from the main frequency where systematics occur.

\section{II. Odd parity microwave interferometer}

\begin{figure}
\begin{center}
\includegraphics[width=3.3in]{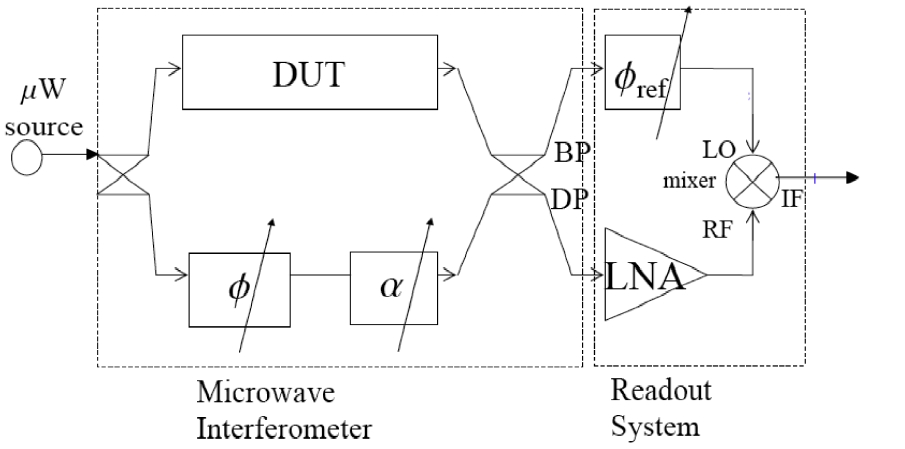}
\caption{The microwave Inerferometer as a precision phase detector. The system consists of the microwave ($\mu W$) pump source, the interferometer, and the phase sensitive readout of the Dark Port (DP). The Bright Port (BP) is used to drive the mixer LO input, and a variable attenuator ($\alpha$) and phase shifter ($\phi$) are used to balance the interferometer.} \label{fig:coaxnms}
\end{center}
\end{figure}

Interferometry is an important precision measurement technology that interferes two waves in and out of phase to create a bright port (BP) and a dark port (DP), where the output of the DP becomes a classical "null detector" and has underpinned a variety of applications over a long period of time \cite{Saulson}. The results presented here are based on the thermal noise limited microwave interferometer \cite{ITW98} shown in Fig. \ref{fig:coaxnms}.  Two systems are presented: 1). A coaxial magnetically asymmetric Mach-Zehnder interferometer system (as proposed in \cite{Tobar}), which is configured as the first ever odd parity rotating test of Lorentz invariance in electrodynamics: 2) The new waveguide power recycled interferometer \cite{Parker,IT09}, which is more sensitive than any laser interferometer operating at the same level of signal power\cite{laser02, laser98} and improves the resolution of phase noise measurements by more than an order of magnitude with respect to current state-of-the-art\cite{IT02}. We show that the new interferometer has the potential to improve the sensitivity of the test of Lorentz Invariance by more than two orders of magnitude in the next generation of experiments.

The magnetic asymmetry of the Mach-Zehnder interferometer (necessary for sensitivity to $\kappa_{tr}$, see \cite{Tobar} for details) is created by introducing in one arm of the interferometer a Device Under Test (DUT) with a different permeability to the balancing arm. In this case the DUT is a string of six magnetically shielded ferrite isolators of effective length 20 cm (length of which the propagating field travels through the ferrite) and relative permeability 0.88 \cite{krupka}. A voltage controlled attenuator and phase shifter located in the other arm maintains balance of the interferometer via a closed loop feedback system of bandwidth 0.25 Hz. The bridge balance can be so exact that the DP is balanced to near zero ($\approx$ -120 dB), so that high gain amplification is useful before the down-mixing process, and overcomes  the relatively high technical fluctuations in the mixing stage. This enables the effective noise temperature of the readout system to be close to its physical temperature \cite{ITW98}.  A stabilized dielectric resonator-oscillator at 9.04 GHz was implemented as the microwave ($\mu W$) source . The amplified signal from the DP is mixed with that from the BP to convert the fluctuations inside the interferometer into voltage noise. The phase of the reference signal ($\phi_{ref}$ in Fig. \ref{fig:coaxnms}) driving the mixer  is adjusted during the calibration procedure to ensure that the voltage noise produced is synchronous with the phase fluctuations of the interferometer.

The interferometer was mounted inside a stainless steel vacuum can on a thermally controlled aluminum plate. Thermal control was necessary to limit phase drift and phase to amplitude conversions. The stainless steel can was located on the rotation table (previously used for a rotating cryogenic oscillator experiment \cite{Stanwix, Stanwix2, Tobar2}), powered via a rotating connector on top of the experiment. The phase variations of the interferometer were inferred from the voltage variations at the output of the mixing stage (IF port in Fig. \ref{fig:coaxnms}) which, in turn, were measured with the digital voltmeter (DVM). Data was collected using a computer that rotated with the experiment. Central to the experiment was keeping track of the interferometer orientation with respect to a universal reference frame. This was achieved in two ways; firstly, by time stamping the DVM measurements with respect to UTC, and; secondly, by triggering the DVM measurements using the orientation of the experiment in the laboratory. The DVM was triggered at a rate of 2 Hz, while the rotation period was approximately 6 seconds, i.e. the DVM was triggered 12 times in one rotation period.

To maintain the balance of the interferometer the output of the mixer was fed back to the voltage controlled phase shifter via a low pass filter of corner frequency 0.25 Hz. The servo suppressed long term fluctuations while the sensitivity of the interferometer at the rotation frequency (0.17 Hz) was maintained at 16.3 v/rad (including filtering effects). From the DVM output we search for periodic signal at the rotation frequency offset by sidereal with the correct phase with respect to the sun centred frame as predicted by the SME \cite{Tobar}. 

\begin{figure}
\begin{center}
\includegraphics[width=3.3in]{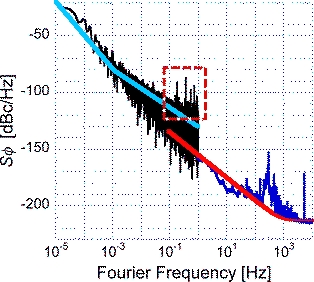}
\caption{Phase noise spectra of the interferometric measurement systems. The noise spectrum of the conventional interferometer (see Fig. \ref{fig:coaxnms}) was measured at Fourier frequencies $f < 1$ Hz. The solid line gives the fit to the noise of stationary experiment. Rotation adds bright lines at the harmonics of the rotation frequency plus some excess broad band noise above 0.1 Hz  (highlighted by the dashed box). The noise spectrum of a power recycled interferometer is displayed at $f > 1$ Hz. No rotation was applied. The interferometer was based on low-loss waveguide components to enhance the efficiency of power recycling and boost its phase sensitivity (see description below). The accompanying solid line is the phase noise model of $1.7\times10^{-8}/f + 2\times10^{-11} rad/\sqrt{Hz} .$} \label{fig:psd}
\end{center}
\end{figure}

\begin{figure}
\begin{center}
\includegraphics[width=3.3in]{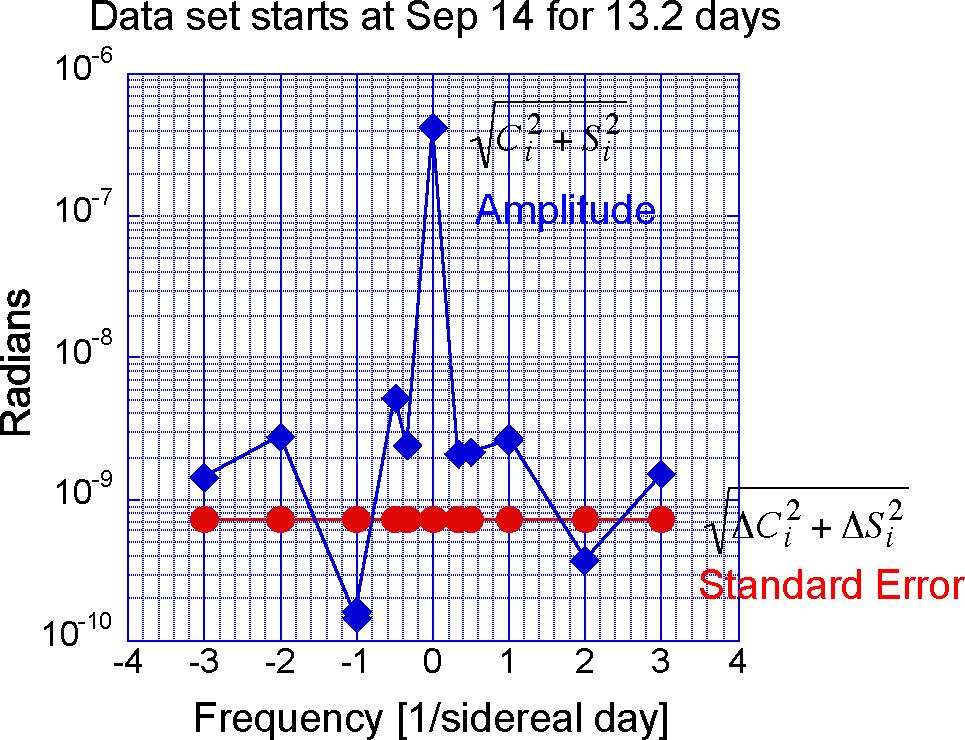}
\caption{Amplitude and standard error for the September 2007 data set as a function of offset frequency from the rotation frequency $\omega_R/2\pi$ in units of $\omega_{\oplus}/2\pi$.} \label{fig:SysLeak}
\end{center}
\end{figure}

Fig. \ref{fig:psd} shows the spectra of phase fluctuations exhibited by the microwave noise measurement systems. Bright lines can be seen that correspond to the rotation frequency and its harmonics. Rotating the experiment also lifted the broadband noise to a level of about -127 dBc/Hz.  For the LI test the noise at 0.17 Hz (the rotation frequency) is the most important. Modulated mechanical vibrations induced by the rotation system were found to be the primary source of systematic in this experiment. The DUT was shielded from stray magnetic fields using nu metal shielding. Oscillating magnetic fields of up to $2.5\times10^{-5}$ Tesla were applied with Helmholtz coils along the x-, y- and z-axis with no response visible.

\section{III. Data Analysis}

To search for a Lorentz violation the data was analyzed using a Demodulated Least Squares (DLS) technique\cite{Stanwix2}, which reduces the size of the data set by performing an initial demodulation of the data with respect to the first harmonic of the rotation frequency ($\omega_R$). Assuming only non-zero $\kappa_{tr}$ in the photon sector of the SME and short data set approximation\cite{Lipa, Stanwix}, i.e. data sets span much less than one year, the phase variations inside interferometer $\Delta\phi$  will be of the form given by Eqs.\ref{nuTest2} to \ref{stage2DAC}:
\begin{equation}
\Delta\phi = A + B t + S(t) ~ {\rm sin}(\omega_{R} t + \varphi) + C(t) ~ {\rm cos}(\omega_{R} t + \varphi) \label{nuTest2}
\end{equation}
\begin{equation}
S(t) = S_0 +  S_{s} ~ {\rm sin}(\omega_\oplus t + \varphi_\oplus)
+  S_{c} ~ {\rm cos}(\omega_\oplus t + \varphi_\oplus)
\label{stage2DAS}
\end{equation}
\begin{equation}
C(t) = C_0 +  C_{s} ~ {\rm sin}(\omega_\oplus t + \varphi_\oplus)
+  C_{c} ~ {\rm cos}(\omega_\oplus t + \varphi_\oplus)
\label{stage2DAC}
\end{equation}
where the coefficients are given by;
\begin{equation}
S_{s}=\frac{4 \pi  L \beta_{0} \kappa_{tr} (\mu_r-1) \cos \eta \cos \chi \cos \Phi_0  }{\lambda }
\label{coefSs}
\end{equation}
\begin{equation}
S_{c}=  \frac{4 \pi  L \beta_{0} \kappa_{tr} (\mu_r-1) \sin  \Phi_0 }{\lambda }
\label{coefSc}
\end{equation}
\begin{equation}
C_{s}= -\frac{4 \pi  L \beta_{0} \kappa_{tr} (\mu_r-1)\cos \chi \sin  \Phi_0 }{\lambda }
\label{coefCs}
\end{equation}
\begin{equation}
C_{c}= \frac{4 \pi  L \beta_{0} \kappa_{tr} (\mu_r-1) \cos \eta  \cos \Phi_0}{\lambda }
\label{coefCc}
\end{equation}
Here, $L$ is the effective length of the interferometer, $\beta_{0}$ is the magnitude of the orbital boost of the Earth, $\mu_r$ is the relative permeability of the magnetic arm of the interferometer, $\lambda$ is the free space wavelength of the microwaves,  $\chi$ is the colatitude (121.8 degrees for Perth), $\eta$ is the angle between the celestial equatorial plane and the ecliptic (23.3 degrees) and $\Phi_{0}$ is the phase of the orbit since the vernal equinox \cite{Tobar2}.

The demodulation was achieved by simultaneously averaging the quadrature amplitudes ($S_{s}, S_{c}, C_{s}, C_{c}$) over a certain number of cycles ($n$) at  $\omega_R$.  Ordinary Least Squares is then applied to the demodulated data set to search for variations at the sidereal frequency $\omega_{\oplus}$. The conversion from radians to $\kappa_{tr}$ \cite{Tobar} for our experiment is given in Tab. \ref{sens}. The optimum number of cycles is determined by the value that gave the minimum standard error, and typically varied between $n$ = 2 to 7 over the individual data sets. At this point the analysis approximates an optimal filter and contributions from fluctuations in the systematic (narrow band noise) will equal the noise contributed by the broad band noise.  

\begin{figure}
\begin{center}
\includegraphics[width=3.3in]{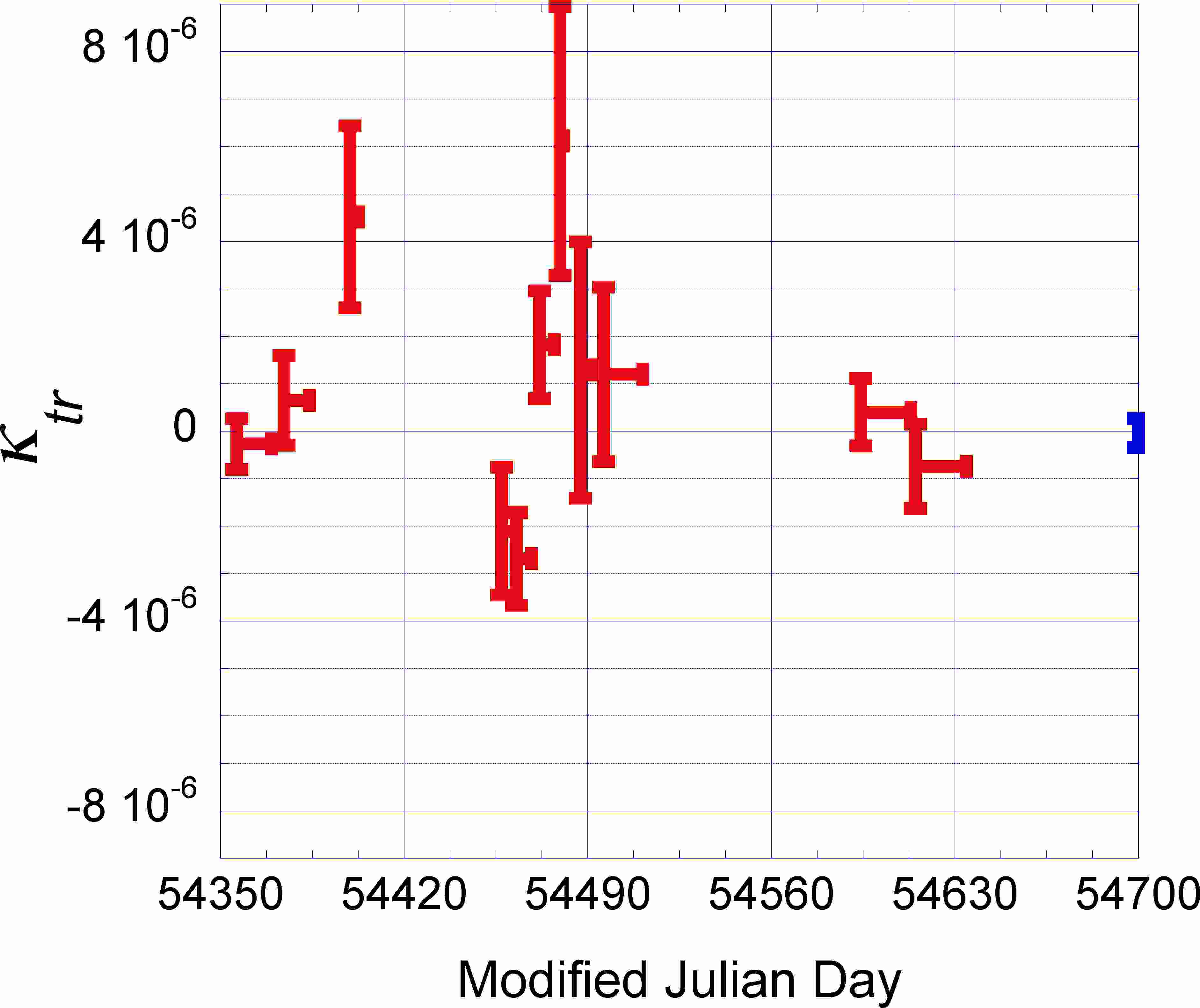}
\caption{The value of $\kappa_{tr}$ from 11 data sets of various lengths from Modified Julian Day 54356 to 54634. The value is extracted from fits to data using DLS at $\omega_R\pm\omega_{\oplus}$. The error bars represent the standard error, and the lengths represent the duration of the data set. In general the longer the duration the smaller the standard error. By taking the weighted average over the multiple data sets the value of $-0.3\pm3\times10^{-7}$ is obtained, which is shown to the right of the figure} \label{fig:kappa}
\end{center}
\end{figure}

Because, the expected Lorentz violating signal with respect to the sun-centered frame occurs at the sidereal offset (from the short data set approximation), we avoid systematic effects in the same way as reference \cite{Stanwix}. Fig. \ref{fig:SysLeak} shows Ordinary Least Squares fit of the amplitude of phase variations (in rads) of the interferometer, as well as  the standard error from a 13.2 day data set during September 2007. The data clearly shows that there is no leakage from the rotational frequency to the sidereal sidebands as the phase amplitudes are not significant. The estimate of $\kappa_{tr}$ as a function of time from 11 data sets are shown in Fig. \ref{fig:kappa}, which gives a final limit of $-0.3\pm3\times10^{-7}$ by calculating the weighted average over all data sets. Other experiments, of the Ives-Stillwell (IS) type are also sensitive to the Robertson Mansouri Sexl (RMS) Lorentz violating time dilation parameter $\alpha$ in the same way as $\kappa_{tr}$ (i.e. the limit put on $\alpha$  is the limit on $\kappa_{tr}$). However, this type of experiment is different in that it is sensitive to $\kappa_{tr}$ but not $\alpha$. Previous IS experiments have put an upper bound on the magnitude of $\kappa_{tr}$ and $\alpha$, of $3\times10^{-8}$ \cite{Carone} , $8.4\times10^{-8}$ \cite{Reinhardt} and $2.2\times10^{-7}$ \cite{Reinhardt, Hoh, Saathoff}.

\begin{table}
\caption{\label{sens} Quadrature amplitudes conversion from output phase to $\kappa_{tr}$ using the short data set approximation. Here $\Phi_{0}$ is the phase of the orbit since the vernal equinox \cite{Tobar2}. The relationship is calculated from the parameters at Perth with respect to the sun centered frame and the experimental parameters of the interferometer\cite{Stanwix, Tobar}.}
\begin{ruledtabular}
\begin{tabular}{ccccc}
Coefficient & Conversion rads to $\kappa_{tr}$ \\
\hline
$S_s$&$4.4\times10^{-4}\cos\Phi_{0}\ \kappa_{tr}$ \\
$S_c$&$-9.0\times10^{-4}\sin\Phi_{0}\ \kappa_{tr}$ \\
$C_s$&$-4.8\times10^{-4}\sin\Phi_{0}\ \kappa_{tr}$ \\
$C_c$&$-8.3\times10^{-4}\cos\Phi_{0}\ \kappa_{tr}$ \\
\end{tabular}
\end{ruledtabular}
\end{table}

\section{IV. Discussion}

Even though the determination of $\kappa_{tr}$ is a factor of 3.5 worse than the current best direct laboratory limit, the experiment is the first of its type and uses only a low power standard interferometer and is vibration limited. The accuracy of phase measurements reported above can be improved by at least two orders of magnitude by providing vibration isolation and switching from the coaxial to the waveguide 'magic tee' based microwave interferometer and making use of a power recycling\cite{Parker, IT09} (note, there is no point switching to the higher sensitive interferometer until the vibration isolation system is installed). Recently, we have built a microwave interferometer where the test sample is placed inside a distributed resonator formed by a short-circuited piece of a waveguide and inductive diaphragm. This extends the interaction time between the test sample and microwave carrier which, in its turn, enhances the random phase/amplitude modulation of the carrier signal by the non-thermal fluctuations in the test sample. The microwave signal reflected from the distributed resonator interferes destructively with a fraction of the incident signal at the dark port of the 'magic tee', which cancels the carrier of the difference signal while preserving the noise modulation sidebands caused by fluctuations in the interferometer arms. In the same way as the coaxial system the noise sidebands are amplifed and demodulated to DC in the non-linear mixing stage resulting in a voltage noise spectral density.

The main reasons for choosing the waveguide components instead of micro-strip (used in \cite{ITW98, IT02}) were $(i)$ to increase the efficiency of power recycling (by reducing the distributed loss in the interferometer arms) and $(ii)$ to minimize the effect of technical noise sources inside the interferometer (micro-strip power splitters could exhibit an excess noise due to poor adhesion of the metal film to dielectric substrate resulting in power-to-phase conversion phenomena). The phase sensitivity of the measurement system was optimized by adjusting the aperture of the inductive diaphragm and its distance from the symmetry plane of the Magic Tee. A piece of hollow waveguide approximately $10 cm$ long was used as the test sample. At $P_{inc} = 1W$ the highest value of phase sensitivity was measured to be $1.4 kV/rad$ with a recycling power enhancement of a factor of 16 (16 W circulating power). 

The phase noise floor of the above measurement system is shown in Fig. \ref{fig:coaxnms}. At Fourier frequencies $f>5kHz$, the phase noise floor was $2\times10^{-11} rad/\sqrt{Hz}$. This is almost an order of magnitude better than the phase resolution of a shot noise limited laser interferometer with power recycling reported in \cite{laser02, laser98}. The above measurements were repeated with the input of the low-noise microwave amplifier terminated to evaluate the contribution of the readout ($LNA$ and $DBM$ assembly) to the overall uncertainty of phase measurements. At low Fourier frequencies the spectral density of phase fluctuations was $1.7\times10^{-8}/f$ $rad/\sqrt{Hz}$, which we believe is a ÒsignatureÓ of ambient temperature fluctuations, which requires further investigation.

\begin{figure}
\begin{center}
\includegraphics[width=3.3in]{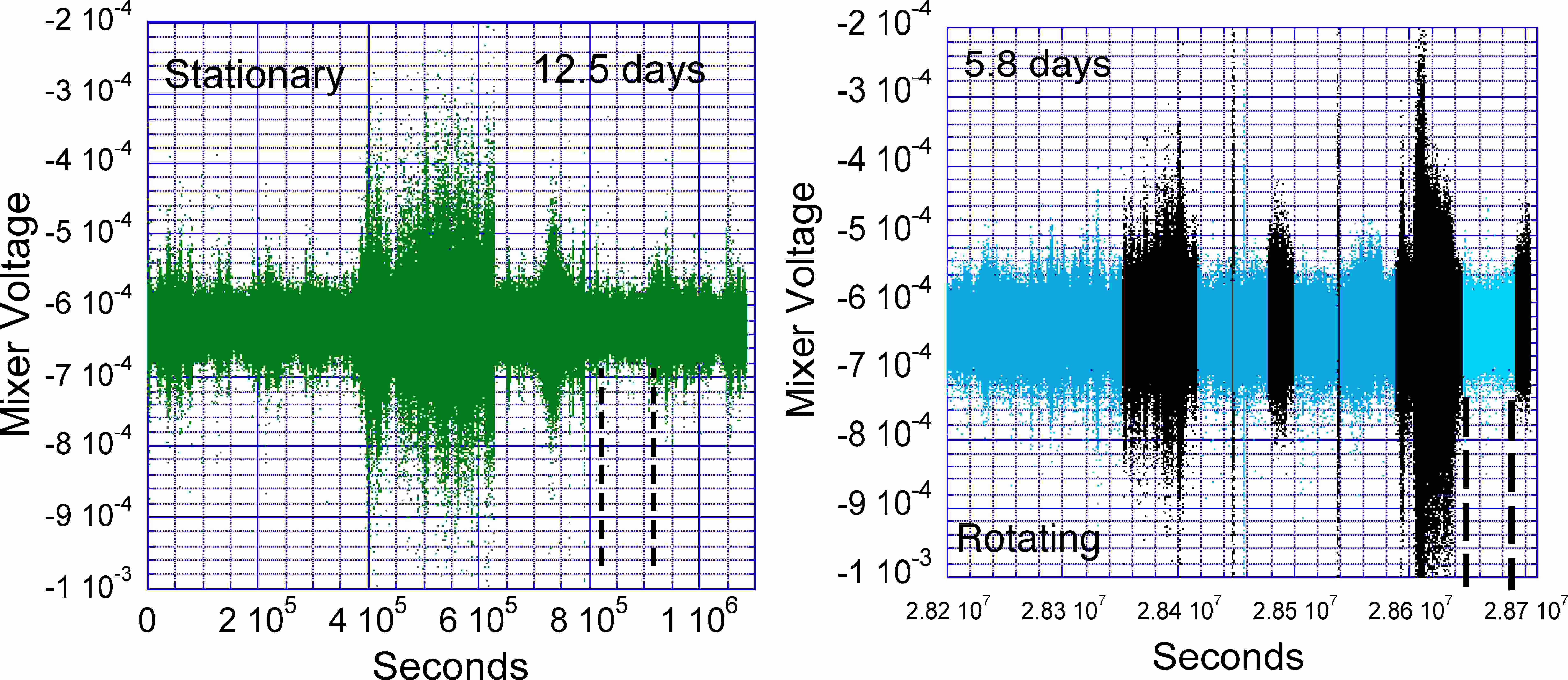}
\caption{Left. The time trace of the mixer output voltage over a 12.5 day period of continuous operation while the experiment was stationary. Intermitent noisy periods are apparent. The phase noise spectral density of the quiet period shown between the dashed lines is shown in fig.\ref{fig:psd}. Right. The time trace of the mixer output voltage over a 5.8 day period of continuous operation while the experiment was rotating. Intermitent noisy periods are apparent (dark regions), which were vetoed in the final analysis. The phase noise spectral density of the quiet period shown between the dashed lines is shown in fig.\ref{fig:psd}.} \label{fig:timetrace}
\end{center}
\end{figure}

Noise measurements with the coaxial interferometer below 1 kHz were sometimes inconsistent showing a large scatter from one experimental run to another. In the best case the noise floor of the coaxial systems was close to that of the waveguide system, while in others it was almost an order of magnitude higher. In this respect, the noise performance of the waveguide based interferometers was always highly reproducible and is one of the reason why the coaxial system had only a 37$\%$ duty cycle as noisy periods were vetoed from the analysis (the system was also out of operation for one month while systematics were investigated). It is not clear, yet, what causes the excess noise in the passive coaxial components, but an example of the noisy periods is compared when the system is stationary and rotating in Fig. \ref{fig:timetrace}.

To calculate the potential sensitivity we use the relationship between radians and phase given in Tab. \ref{sens}. Given that we use all four coefficients to determine the standard error of $\kappa_{tr}$ ($\delta\kappa_{tr}$), then we can write the following simple relation to approximate the sensitivity.
\begin{eqnarray}
\delta\kappa_{tr} \approx  \frac{1}{10^{-3}}\frac {\delta\phi}{\sqrt{f_{R}\tau_{obs}}}
\label{eq:sens}
\end{eqnarray}
Here $\delta\phi$ is the spectral density of $rms$ phase fluctuations in $rads/\sqrt{Hz}$, $\tau_{obs}$ is the total observation time in seconds and $f_{R}$ is the rotation frequency in Hz. The phase noise spectral density at 0.17 Hz was $10^{-7}$ $rads/\sqrt{Hz}$ (-140 dBc/Hz) for the recycled waveguide interferometer. Thus, for the parameters of our experiment, we have the possibility of determining $\kappa_{tr}$ to $5\times10^{-8}$ using the recycled interferometer. Faster rotation frequencies could also be used to improve performance. For example, by simply increasing the rotation rate by a factor of ten (1.7 Hz) the phase noise  would be $10^{-8}$ $rads/\sqrt{Hz}$ (-160 dBc/Hz) and in the same amount of time a sensitivity of $10^{-9}$ could be achieved. It is also important to identify the source of low frequency$1/f$ phase fluctuations and of ways to reduce them. If one could achieve the Nyquist thermal noise limit, given by $2\times10^{-11}$ $rads/\sqrt{Hz}$, a sensitivity of order $10^{-12}$ could be achieved. 

The recycled interferometer is also ideally suited for studying the noise phenomena in low-loss components and materials at microwave frequencies. We have used it to characterize the intrinsic phase fluctuations in ferrite circulators (used for this test). So far, we could only claim that there is no general rule describing the phase noise in such devices: in some cases the circulator phase fluctuations were easily observable, while in others no noise was detected. For the rotating experiment we chose isolators that exhibited no measurable phase noise. We also constructed a ferrite loaded waveguide impedance-matched on both ends, which exhibited no measurable noise. This will be used as a DUT in our future experiment based on the waveguide interferometer placed on top of the vibration isolated platform.

Finally we mention that a model dependent theory-based analysis has recently provided improved indirect bounds by observing gamma rays from cosmic sources \cite{Klink}. Also, new high energy experiments, which are non-sensitive to the odd parity terms have also set new bounds on $\kappa_{tr}$ \cite{Altschul, MH1, MH2}. These bounds are tighter, however are set at a very different energy scale.

\begin{acknowledgments}
The authorÕs thank Peter Wolf, Alison Fowler and Michael Miao for their assistance. This work was funded by the Australian Research Council.
\end{acknowledgments}

\end{document}